\documentstyle[12pt] {article}
\newcommand{\beq}{\begin{equation}}
\newcommand{\eeq}{\end{equation}}
\newcommand{\beqa}{\begin{eqnarray}}
\newcommand{\eeqa}{\end{eqnarray}}
\newcommand{\ba}{\begin{array}}
\newcommand{\ea}{\end{array}}

\begin{document}
\baselineskip 24 true pt

\def \mtil{{\tilde m}}

\begin{center}
{\bf HALF-LIVES AND PRE-SUPERNOVA WEAK INTERACTION \\
RATES FOR NUCLEI AWAY FROM THE STABILITY LINE} \\
\end{center}
\vskip 0.5 truecm
\begin{center}
{ K.Kar$^{1}$, S.Chakravarti$^{2}$, A.Ray$^{3,4,5}$ and S.Sarkar$^{6}$ \\ \ \\
\vskip 0.25 truecm

\hoffset =-.5 truecm
$^1$ Saha Institute of Nuclear Physics, 1/AF Bidhannagar, Calcutta 700064, India \\
$^2$ Physics Department, Visva Bharati University, Santiniketan 731235, India \\

$^3$ Tata Institute of Fundamental Research, Homi Bhabha Road, Mumbai 400005, India \\

$^4$ Institute of Theoretical Physics, University of California, Santa Barbara, CA93106, USA \\
$^5$ NASA/Goddard Space Flight Center,Code 661, Greenbelt, MD20771, USA \\

$^6$ Ananda Mohan College, 102/1 Amherst Street, Calcutta 700009, India \\}

\end{center}
\textwidth=13.0 truecm
\vskip 0.5 truecm

{\bf Abstract:} A detailed model for the calculation of beta decay rates of
the $fp$ shell nuclei for situations prevailing in pre-supernova and
collapse phases of evolution of the core of massive stars leading to
supernova explosion has been extended for electron-capture rates.
It can also be used  to determine the half-lives of neutron-rich nuclei in 
the $fp/fpg$ shell.  The model uses an averaged Gamow-Teller (GT) strength
function. But it can  also use the experimental log ft values 
and GT strength function from $(n,p)$ reaction studies wherever available. 
The calculated  rate includes contributions  from each of the low-lying excited
states of the mother including some specific resonant states ("back resonance")
having large GT matrix elements.

\section{Introduction}

Nuclei away from the line of stability play an important role in a number of
problems in nuclear astrophysics. The wealth of experimental information on
such nuclei right up to the drip line ones
obtained from  experiments with
radioactive nuclear beams (RNB) in recent years has focussed many issues in
the r-process nucleosynthesis, stellar evolution, etc. In this article we
discuss why the weak interaction rates of these nuclei in stellar
conditions are important for the pre-supernova (preSN) and supernova (SN)
evolution of massive stars and describe our work in constructing a
model for calculating these rates for nuclei important at this stage with $50<A<80$.
This model also predicts the half-lives of neutron/proton-rich nuclei, an
important ingredient in the structure studies of the RNB nuclei. The beta-decay
of the n-rich nuclei plays a significant role at the silicon burning stage of the preSN,
and the rates for some nuclei considered important at typical temperatures and densities
have been reported [1].
The electron capture (EC) rates on nuclei with neutrons filling the fp-shell
orbits are also needed for the collapse state of the SN. We report here on
the model for the calculation of the EC rates for them.

\section{Pre-Supernova and Supernova Evolution}

The last stage of stellar burning involves successive addition of
$\alpha$-particles
on silicon and other intermediate mass nuclei at temperatures typically at
$4\times 10^9$ K. In the time scale of a few days the final end product $^{56}$Fe
is reached and a degenerate core is created where no further energy
generation through nuclear reactions is possible. If the core mass exceeds the
Chadrasekhar mass then the gravitational collapse of the core starts with the
matter in the envelope having a quasi-free fall [2]. During the further
evolution of the core one of the most important quatities is the lepton fraction
 $Y_l$. This is because
at densities a few times the nuclear matter density the collapse ultimately
stops with the stiffer baryonic matter at the central region giving rise to a
shock wave whose energy is strongly dependent on $Y_l$, being proportional to
$Y_{lf}^{10/3}(Y_{lf}-Y_{li})$,
where $Y_{lf}$ and $Y_{li}$ are the final and initial lepton fractions. Recent
findings indicate that even at the Si-burning stage beta decay and
electron captures on the relevant nuclei are important in order to determine
the $Y_e$ (the electron fraction) at the beginning of the collapse and the
network of nuclei involved in the weak interaction processes needs to be
enlarged with the inclusion of some neutron-rich nuclei [3,4]. For this purpose
beta decay rates on some nuclei in the fp shell have recently been calculated
[1,5] and the possible set of important nuclei at different densities and
temperatures have been listed with the simplified assumption of nuclear
statistical equilibrium. Here one finds that some neutron-rich nuclei
contribute significantly to leptonisation
in spite of their small abundances
since they have large beta decay rates because
of large Q values. Electron capture rates on some of these fp shell nuclei
again become very important during the collapse phase. During the collapse
deleptonisation takes place through rapid electron capture on nuclei with
$A>56$ until the neutron number exceeds 40  when the allowed electron capture
gets blocked.

\bigskip

\section{Half-Lives and Beta-decay Rates of Neutron - rich Nuclei}

 We first briefly describe the physics input in the model to calculate the
presupernova beta-decay rates [1]. It is based on a statistically averaged
allowed beta-decay, in particular the Gamow-Teller(GT) strength function.
For the GT sum rule strength for $\beta^-$ decay we use an expression
involving the occupancies of neutron particles and proton holes which is seen
to be a good approximation for the fp shell nuclei using the framework of
spectral distribution theory [6,7,8].
One observes that for $\beta^-$ decay with $N>Z$,
one can reach only the tail of the GT giant resonance due to energy
conservation. This giant resonance is assumed to be distributed as a Gaussian
in energy, a feature seen in studies in large full shell-model spaces in the
absence of strong collectivity. The centroid of this Gaussian is fixed from
the results of (p,n) reaction studies and a phenomenological formula for its
position involving $(N-Z)$ and $A^{1/3}$ is employed [9]. Recently, Sutaria and
Ray [10] improved this by using a fit to new data with mostly fp shell nuclei
and a calculation of rates based on their centroids is being persued.
The strength width coming from the nuclear Hamiltonian is left as a parameter
and is determined globally by a best fit to the observed half-lives of a
number of n-rich fp shell nuclei with $A>60$. It is seen that large shell
configuration mixing and possible coupling to $\Delta$ excitation gives rise
to a long tail in the GT strength distribution at higher energies causing
quenching of the observed strength. This is taken into account by an overall
quenching factor of 0.6 in the sum rule following Aufderheide {\it et al.} [4].
In this work we report the calculated half-lives of some more fp shell nuclei
with neutrons filling the $g_{9/2}$ orbit. For the latter case we put the
neutrons in excess of 40 in the $g_{9/2}$ orbit and add three times the number of
neutrons in the $g_{9/2}$ orbit to the GT sum rule. For example for $^{68}$Fe,
we put 2 neutrons in $g_{9/2}$ and add 6 to the GT sum rule calculated for
$^{66}$Fe. In Table 1 we present our predictions compared with experimental
values as well as with predictions by microscopic theories [11] and gross theory
or its modified form [12]. In Table 2, column A is with a Gaussian GT
distribution with the width parameter set at 6.3 MeV and column B is with the same
width at 7.5 MeV and a skewed distribution with skewness -0.3 to bring more
strength in the ground state domain.
In the calculation of rates in stellar conditions the partial blocking of the
electrons produced by the Fermi sea outside is important and is taken into account
using the electron chemical potential obtained by solving an integral equation
numerically.
This chemical potential has built into it both finite temperature correction and
corrections coming from the production of electron-positron pairs . The model also
includes contributions from excited states which, though reduced by the factor
$exp(-E_{i}/kT)$, are appreciable because of larger Q-values than
the ground state. For their sum rule strength we use the same expressions
where the occupancies are calculated by spectral distributions at the respective
excitation energies and the strength distribution is assumed to have the same form.
For some nuclei at high densities ($\rho>10^9 g cm^{-3}$) these excited state contributions
are an order of magnitude more than the ground state contributions.

In Table 2 we present the $\beta^-$ decay rates for typical densities and temperatures
for four isotopes of cobalt alongwith their Q-values. We find that the rates
are the highest for the largest Q-value for $^{64}$Co. Our rates for some
nuclei include experimental log ft's to the ground state/lowlying excited state
where the predicted half-lives without them are in disagreement. For $^{62}$Co and $^{62}$Fe one
finds that this inclusion of one or two lowlying log ft's brings the predicted
half-lives very close to experimental values. The rates also include
contributions from a particular resonant state of the mother nucleus which is in
equilibrium with the electron capture rate on the daughter. This resonant state
has a large overlap with
the daughter ground state or low excited state and can get connected to them by proton-neutron
single particle
allowed GT transition.
These so called
`back resonance' rates are calculated by following the prescription of Fuller, Fowler
and Newman II [13] and is described in Kar, Ray and  Sarkar [1].

\bigskip

\section{Model for Electron Capture Rates}

We briefly describe the major features of the calculation for the electron
capture rates. They are:

1) The GT sum-rule for electron capture uses the same expressions as in the
case of beta-decay, replacing the neutron occupancy by the proton occupancy and
the proton hole occupancy by the neutron hole occupancy.
The quenching factor used is 0.5, a typical value from (n,p) data.
Thus the GT sum for EC ($S_{\beta^+}$) is much smaller than the corresponding
$\beta^-$ sum-rule ($S_{\beta^-}$) on the same nucleus and the difference between
the two is always $3(N-Z)$, a model independent result. For example, for $^{62}$Fe
the unquenched $S_{\beta^-}$ and $S_{\beta^+}$ are 36.7 and 6.7 whereas the
corresponding quenched values are 21.5 and 3.4 respectively.

2) For the strength distribution one uses the Gaussian form in the case of EC rates
also. For the centroid of the Gaussian one uses the empirical form obtained
by Sutaria and Ray [10] through fits to (n,p) data and for the width one uses
an average value of 2 MeV.

3) For nuclei with $N>Z$, EC connects the ground state of the mother nucleus with
isospin $T_0$ to states of the daughter nucleus with isospin $(T_0+1)$ only. As
a result, Fermi transitions are ruled out by isospin conservation.

4) The phase space factor for EC is of course different from the $\beta^-$ decay
with 2 particles in the exit channel instead of 3.

The code used for this calculation has two sectors with electron capture
Q-values both positive and negative.
For the positive Q-value case, for the ground state
one has the options of using (i) experimental log ft's to low-lying states of the
daughter, (ii) shell model log ft's,  and (iii) GT strength function of our model.
For the negative Q-value case the code can calculate the ground state rates from
(i) B$_{GT}$(E) distribution of (n,p) reactions given as strength/MeV;
(ii) shell model log ft's and (iii) GT strength-function of our model.
For the low-lying excited states in both the cases the only options available are
using (i) shell model log ft's, if available, and
(ii) the GT strength function of our model. For electron capture
a special excited state of the mother is considered whose shell model
configuration differs from the ground state of the daughter by a single proton and
a single neutron so that it has a large overlap with the daughter state and the
electron capture rate is in equilibrium with the back reaction rate of the
daughter. In Table 3 we give EC rates on $^{56}$Fe for typical preSN temperatures
and densities. One sees that the contributions from the excited states are sometimes
an  order of magnitude larger than the rate from the ground state of the nucleus.
For $^{56}$Fe the values with 28 excited states are compared to
the contribution coming from the lowest 4 excited states.

\section{Conclusion}

In conclusion, we stress that we have built models for calculating the beta decay
and electron capture rates for fp shell nuclei for use in problems of
preSN and SN evolution. This model can be used to predict half-lives of
beta decay of nuclei away from the line of stability. We plan to extend
this to predict half-lives for very n-rich nuclei near the drip line as well
as to electron capture and positron decay for the p-rich nuclei. This also should
be extended to lower densities and temperatures so that it can be used to
supply the weak interaction rates in the r- and s-process nucleosynthesis calculations.

\bigskip

\noindent{\bf References}

[1] Kar K, Ray A and Sarkar S 1994 {\it Ap.J.} {\bf 434} 662

[2] Bethe H A, Brown G E, Applegate J and Lattimer J 1979 {\it Nucl.
    Phys. A} {\bf 234} 487

[3] Bethe H A 1990 {\it Rev. Mod. Phys.}{\bf 62} 801

[4] Aufderheide M B, Brown G E, Kuo T T S, Stout D B and Vogel P
    1990 {\it Ap.J.} {\bf 362} 241

[5] Aufderheide M B, Fushiki I, Woosley S E and Hartmann D H, 1994 {\it Ap. J.
    Suppl.}{\bf 91} 1389

[6] Draayer J P, French J B and Wong S S M, {\it Ann.Phys. (N.Y.)} 1997
    {\bf 106} 472; 1997 {\bf 106} 503

[7] Kota V K B and Kar K 1988 Univ. of Rochester, Laboratory Report No
    UR-1058; 1989 {\it Pramana} {\bf 32} 647

[8] French J B and Kota V K B  1982 {\it Annual Rev. Nucl Part. Sci.}
    {\bf 32} 35

[9] Nakayama K,  Pio Galeao a and Krmpotic F, {\it Phys. Lett.} 1982
    {\bf 114} 217

[10] Sutaria F K and Ray A, 1995 {\it Phys. Rev.} {\bf C52} 3460

[11] Klapdor H V, Metzinger J and Oda T, {\it Atomic Data and Nucl. Data Tables}
     1984 {\bf 31} 81

[12] Takahashi K, Yamada J M and Kondoh J T ,{\it  Atomic Data and Nucl. Data Tables} 1973 {\bf 12} 101

[13] Fuller G M, Fowler W A and Newman M J, {\it Ap.J.} 1982 {\bf 252} 715

\vfil\hfil\eject

\centerline {{\bf Table 1}}
{\bf{Comparison of the $\beta^{-}$ decay rates from the ground
state of the mother nucleus for the four different isotopes of Cobalt.
Values are given for $Y_e = 0.47$.}}
$$\vbox{\halign{&\hfil  #  \hfil \cr
\noalign{\hrule}\cr
\noalign{\hrule}\cr
&& \multispan{4}\hfil Nucleus \& Q-value (MeV)\hfil\cr
&Temperature &\multispan{4}\hrulefill\cr
 log $\rho_{10}$ &(K) &$^{62}$Co&$^{63}$Co&$^{64}$Co&$^{65}$Co\cr
&&5.322&3.674&7.307&5.970\cr
\cr
\noalign{\hrule}\cr
& 3 $\times$ 10$^9$~& 4.92 $\times$ 10$^{-3}$ ~& 6.29 $\times$ 10$^{-3}$~&
6.32 $\times$ 10$^{-1}$~& 9.01 $\times$ 10$^{-2}$\cr
-2.0 & 4 $\times$ 10$^9$~& 5.12 $\times$ 10$^{-3}$ ~& 6.88 $\times$ 10$^{-3}$~&
6.39 $\times$ 10$^{-1}$~& 9.24 $\times$ 10$^{-2}$\cr
& 5 $\times$ 10$^9$~& 5.33 $\times$ 10$^{-3}$ ~& 7.49 $\times$ 10$^{-3}$~&
6.47 $\times$ 10$^{-1}$~& 9.53 $\times$ 10$^{-2}$\cr
\noalign{\hrule}\strut\cr}}$$
\vfil\hfil\eject
\bigskip
\bigskip
\centerline {{\bf Table 2}}
\centerline{{\bf {Comparison of calculated and experimental Half-Lives.}}}
$$\vbox{\halign{&\hfil\  # \ \hfil \cr
\noalign{\hrule}\cr
\noalign{\hrule}\cr
&& \multispan{4}\hfil Calculated Half-life $\tau_{1/2}$ (s)\hfil\cr
&&\multispan{4}\hrulefill\cr
 Nucleus &Experimental&\multispan {2} \hfil This work \hfil&Gross Theory&QRPA\cr
&&                     \multispan {2} \hrulefill\cr
&&                     A &B \cr
\cr
\noalign{\hrule}\cr
$ ^{68}Co$&  0.18$\pm$0.10  & 0.54& 0.10   & 0.81 & 0.80 & \cr
$ ^{68}Fe$&  0.10$\pm$0.06  & 1.05& 0.36   & 0.42 & 0.37 & \cr
$ ^{69}Ni$&  11.4           & 6.53& 2.46   & 20 & 19.9 & \cr
$ ^{71}Cu$&  19.53          & 13.12& 5.25  & 7.6 & 35 & \cr
$ ^{72}Cu$&  6.6            & 1.47& 0.38   & 2.7 & 7.0 & \cr
$ ^{73}Cu$&  3.9            & 2.75& 0.96   & 1.7 & 1.0 & \cr
$ ^{69}Co$&  0.27           & 0.72& 0.19   & 0.68 & 0.73 & \cr
$ ^{71}Ni$&  1.86           & 1.85& 0.63   & ~~ & ~~ & \cr
$ ^{72}Ni$&  2.10           & 4.25& 1.87   & ~~ & ~~ & \cr
$ ^{58}Cr$&  7.0            & 22.0& 10.0   & ~~ & ~~ & \cr
$ ^{63}Mn$&  0.25           & 0.94& 0.26   & ~~ & ~~ & \cr
$ ^{59}Cr$&  0.74           & 1.81& 0.54   & ~~ & ~~ & \cr
\noalign{\hrule}\strut\cr}}$$
\vfil\hfil\eject
\bigskip
\bigskip
\centerline {{\bf Table 3}}
\medskip
{{\bf{ Electron-capture rates for $ ^{56}Fe$.
Values are given for $Y_e = 0.47$ and
$ T = 6 \times 10^9$K.}}}
$$\vbox{\halign{&\hfil  # \hfil \cr
\noalign{\hrule}\cr
\noalign{\hrule}\cr
log $\rho_{10}$ &$\mu_e$ in MeV  &Rate (g.s.)&Rate & Rate
& Total Rate \cr
&&(logft=2.48)&with 4 states& with 28 states &~~~~~&\cr
\noalign{\hrule}\cr
-2.0 & 1.00 & 2.4 $\times$ 10$^{-4}$ & 1.46 $\times$ 10$^{-3}$&
4.39 $\times$ 10$^{-3}$ & 4.39 $\times$ 10$^{-3}$\cr
-1.0 & 3.40 & 0.008 & 0.224 & 0.396 & 0.396 \cr
\noalign{\hrule}\strut\cr}}$$
\vfil\hfil\eject

\end{document}